\documentstyle[twoside,fleqn,espcrc2]{article}
\newcommand{\cL}{{\cal L}}

\newcommand{\Paris}{Par\'\i{}s}

\newcommand{\beq}{\begin{equation}}
\newcommand{\eeq}{\end{equation}}
\newcommand{\beqn}{\begin{eqnarray}}
\newcommand{\eeqn}{\end{eqnarray}}

\newcommand{\cp}{c_{\mathrm photon}}

\newcommand{\cn}{c_{\mathrm gravity}}

\newcommand{\gem}{g_{\mathrm em}}

\newcommand{\AmS}{{\protect\the\textfont2
  A\kern-.1667em\lower.5ex\hbox{M}\kern-.125emS}}
\title{$\chi$Variable-Speed-of-Light Cosmologies
} 
\author{S. Liberati\address{International School for Advanced Studies 
         (SISSA), Via Beirut 2--4, 34014 Trieste, Italy}
         \thanks{Istituto Nazionale di Fisica Nucleare (INFN),
         sezione di Trieste, Italy},
	B.A. Bassett\address{Relativity and Cosmology Group,
         Portsmouth University,~PO1~2EG, UK}
	 \thanks{Department of Theoretical Physics,
         University of Oxford, 1 Keble Road, OX1 3NP, UK},
        C. Molina--\Paris\address{Theoretical Division, Los Alamos
         National Laboratory, Los Alamos, New Mexico 87545, USA}, and 
	Matt Visser\address{Physics Department, Washington University,
         Saint Louis, Missouri 63130-4899, USA}}
\begin{document}
\begin{abstract}
Variable-Speed-of-Light (VSL) cosmologies are currently attracting
much interest as a possible alternative to cosmological inflation.  We
discuss the fundamental geometrodynamic aspects of VSL cosmologies,
and provide several alternative implementations.  These
implementations provide a large class of VSL cosmologies that pass the
zeroth-order consistency tests of being compatible with both classical
Einstein gravity and low-energy particle physics. While they solve the
``kinematic'' puzzles as well as inflation does, VSL cosmologies
typically do not solve the flatness problem since in their purest form
no violation of the strong energy condition occurs. Nevertheless,
these models are easy to unify with inflation.
\end{abstract}

\maketitle

\section{INTRODUCTION}
Variable-Speed-of-Light (VSL) cosmologies have recently generated
considerable interest as alternatives to the inflationary
framework. They serve both to sharpen our ideas concerning
falsifiability of the standard inflationary paradigm, and also to
provide a contrasting scenario that is itself amenable to
observational test.  In this presentation we wish to assess the
internal consistency of the VSL framework, and ask to what extent it is
compatible with geometrodynamics (Einstein gravity). This will
lead us to propose a particular class of VSL models that implement
this idea in such a way as to inflict minimal ``violence'' on GR, and
which at the same time are ``natural'' in the context of one-loop
QED. For a detailed discussion of all these issues we refer to the
paper \cite{BLMV99} which has inspired the present talk.

The question of the intrinsic compatibility of VSL models with GR is
not a trivial one: Ordinary Einstein gravity has the constancy of the
speed of light built into it at a deep and fundamental level; $c$ is
the ``conversion constant'' that relates time to space. Even at the
level of coordinates we need to use $c$ to relate the zeroth
coordinate to Newtonian time: $d x^0 = c \;
dt$. Thus, simply replacing the {\em constant} $c$ by a
position-dependent {\em variable} $c(t,\vec x)$, and writing $dx^0 =
c(t,\vec x)\; dt$ is a highly suspect proposition.

If this substitution is performed at the level of the metric, it is
difficult to distinguish VSL from a mere coordinate change (under
such circumstances VSL has no physical impact). Apparently more
attractive, (because it at least has observable consequences), is the
possibility of replacing $c \to c(t)$ directly {\em in the Einstein
tensor} itself.
[This is the route chosen by
Barrow--Magueijo~\cite{Barrow98a,Barrow98b,Barrow98c}, by
Albrecht--Magueijo~\cite{Albrecht98}, and by Avelino--Martins
\cite{Avelino}.]
If one does so, the modified ``Einstein tensor'' is  
{\em not} covariantly conserved (it does {\em not} satisfy the
contracted Bianchi identities), and this modified ``Einstein tensor''
is not obtainable from the curvature tensor of {\em any} spacetime   
metric.  Indeed, if we define a timelike vector $V^\mu = (\partial/
\partial t)^\mu = (1,0,0,0)$ then a brief computation yields
\begin{equation}
\nabla_\mu G_{\mathrm modified}^{\mu\nu} \propto \dot c(t) \; V^\nu.
\end{equation}
Thus violations of the Bianchi identities for this modified ``Einstein
tensor'' are part and parcel of this particular way of trying to make
the speed of light variable.  If one now couples this modified
``Einstein tensor'' to the stress-energy via the Einstein equations,
then the stress-energy tensor (divided by $c^4$) cannot be covariantly
conserved either, and so it cannot be variationally
obtained from {\em any} action.
To our minds, if one really wants to say that it is the speed of light
that is varying, then one should seek a theory that contains two
natural speed parameters, call them $\cp$ and $\cn$, and then ask that
the ratio of these two speeds is a time-dependent (and possibly
position-dependent) quantity.  To implement this idea, it is simplest
to take $\cn$ to be fixed and position-independent. 
So doing, $c_{\mathrm gravity}$ can be safely used in the usual way to set
up all the mathematical structures of differential geometry needed in
implementing Einstein gravity. 
 
\section{TWO--METRIC VSL COSMOLOGIES}
Based on the preceding discussion, we feel that the first step towards
making a VSL cosmology ``geometrically sensible'' is to write a {\em
two-metric} theory in a form where the photons couple to a second
electromagnetic metric, distinct from the spacetime metric that
describes the gravitational field. 
[Somewhat related two-metric implementations of VSL cosmology are
discussed by Clayton and Moffat~\cite{Clayton-Moffat-1,Clayton-Moffat-2}
and by Drummond~\cite{Drummond}. See~\cite{BLMV99} for details on how
those implementations differ from our own.] 
This permits a precise physical meaning for VSL: If the two null-cones
(defined by $g$ and $\gem$, respectively) do not coincide at some time,
one has a VSL cosmology. 

We want to stress that the basic idea of a quantum-physics-induced
effective metric, differing from the spacetime metric (gravity metric),
and affecting only photons is actually far from being a radical point of
view. This concept has gained in the
last decade a central role in the discussion of the propagation of
photons in non-linear electrodynamics. In particular we stress that
``anomalous'' (larger than $\cn$) photon speeds have been calculated
in relation with the propagation of light in the Casimir vacuum
\cite{Sch90}, as well as in gravitational fields \cite{DH80}.

Within our own framework, alternative approaches can be (I) 
to couple just the photons to $\gem$ meanwhile all the other matter and
gravity couple to $g$, or (II) to couple all the gauge bosons to $\gem$,
but couple everything else to $g$, or (III) to couple {\em all} the matter
fields to $\gem$, keeping gravity as the only field coupled to $g$.  
A particularly simple EM metric is
\begin{equation}   
[\gem]_{\alpha\beta} =  
g_{\alpha\beta} -
(A\; M^{-4}) \;
\nabla_\alpha \chi \; \nabla_\beta \chi.
\end{equation}
Here we have introduced a dimensionless coupling $A$, and taken
$\hbar=\cn=1$, in order to give the scalar field $\chi$ its canonical
dimensions of mass-energy.  The normalization energy scale, $M$, is
defined in terms of $\hbar$, $G_{\mathrm Newton}$, and $\cn$.
Provided $M$ satisfies $M_{\mathrm Electroweak}<M<M_{\mathrm Pl}$, the EM
lightcones can be much wider than the standard (gravity) lightcones
without inducing a large backreaction on the spacetime geometry due to
the scalar field $\chi$. The presence of this dimensionfull
constant implies that $\chi$VSL models will automatically be
non-renormalizable. $M$ is then the energy at which the
non-renormalizability of the $\chi$ field becomes important.  
So these models should be viewed as ``effective
field theories'' valid for sub-$M$ energies. In this regard
$\chi$VSL implementations are certainly no worse behaved than many of
the models of cosmological inflation and/or particle physics currently
extant.

The evolution of the scalar field $\chi$ will be assumed to be
governed by some VSL action $S_{\mathrm VSL}$. Then the complete
action for the first of the models proposed above is
\begin{eqnarray}
S_{I} &=&
\int d^4 x \sqrt{-g} \; \left[ R(g) + \cL_{\mathrm matter} \right] 
\nonumber\\
&+&
\int d^4 x \sqrt{-\gem} \;
\gem^{\alpha\beta}  \; F_{\beta\gamma} \;
\gem^{\gamma\delta} \; F_{\delta\alpha}
\nonumber\\
&+&
\int d^4 x \sqrt{-g} \; \cL_{\mathrm VSL}(\chi).
\end{eqnarray}

Let us suppose the potential in this VSL action has a global minimum, but
that the $\chi$ field is displaced from this minimum in the early
universe: either trapped in a meta-stable state by high-termperature
effects or displaced due to chaotic initial conditions. The transition to
the global minimum may be either of first or second order and during it 
$\nabla_\alpha \chi \neq0$, so that $\gem \neq g$. Once the
true global minimum is achieved, $\gem = g$ again. Since one can arrange
$\chi$ today to have settled to the true global minimum, current
laboratory experiments would automatically give $\gem=g$. 

Since $V(\chi)\geq 0$ in the early universe, $\chi$ could drive an
inflationary phase.  While this is true we stress instead the more
interesting possibility that, by coupling an independent inflation field
to $\gem$,  $\chi$VSL models can be used to improve the inflationary
framework by enhancing its ability to solve the cosmological
puzzles~\cite{BLMV99}.

\section{COSMOLOGICAL PUZZLES}

The general covariance of General Relativity means that the set of
models consistent with the existence of the apparently universal class
of preferred rest frames defined by the Cosmic Microwave Background
(CMB) is very small and non-generic. Inflation seeks to alleviate this
problem by making the flat Friedmann--Lemaitre--Robertson--Walker
(FLRW) model an attractor within the set of almost--FLRW models, at
the cost of violating the strong energy condition (SEC) during the
inflationary epoch. VSL cosmologies by contrast typically sacrifice
Lorentz invariance, again thereby making the flat
FLRW model an attractor \cite{Barrow98a,Barrow98b,Barrow98c,Albrecht98}.

Our own approach, while is able to solve the ``kinematic'' puzzles as
well as inflation does, cannot solve the flatness problem since in its
purest formulation (no inflation driven or enhanced by $\chi$) violations
of the SEC do not occur, and because our models do not lead to an explicit
``hard'' breaking of the Lorentz invariance like other VSL models do.
[Our class of VSL models exhibit a ``soft'' breaking of Lorentz
invariance,  which is qualitatively similar to the notion of spontaneous
symmetry breaking in particle physics.]

We will now consider some of the major cosmological puzzles,
directing the reader to \cite{BLMV99} for an extended discussion.

\subsection{Isotropy}

One of the major puzzles of the standard cosmological model is that
the isotropy of the CMB seems in conflict with estimates of the region
of causal contact at last scattering.  The basic mechanism by which
VSL models solve this cosmological puzzle relies on the fact that the
(coordinate) size of the horizon at the time of last scattering $t_*$
is modified by the time dependence of the photon speed $R_{\mathrm hz}
= \int_0^{t_*} {\cp \; dt / a(t)}$.  It is this quantity that sets the
distance scale over which photons can transport energy and thermalize
the primordial fireball.  On the other hand, the coordinate distance
out to the surface of last scattering is $R_{\mathrm ls} =
\int_{t_*}^{t_0} {\cp \; dt / a(t)}$.  Observationally, the
large-scale homogeneity of the CMB implies $R_{\mathrm hz} \geq
R_{\mathrm ls}$. Although this is a paradox in the standard
cosmological framework (without inflation), it can be achieved by
having $\cp \gg \cn$ early in the expansion and keeping $\cp \approx
\cn$ between last scattering and the present epoch (as it should be
for VSL models to be compatible with observations at low-redshift).

\subsection{Flatness}

The flatness paradox arises from the fact that the flat FLRW universe,
although plausible from observation, appears as an unstable solution
of GR.  {From} the Friedmann equation, it is a simple matter of
definition that 
\begin{equation}
\epsilon\equiv\Omega-1=
\frac{K\;c^2}{H^{2}\;a^{2}}
=\frac{K \;c^{2}}{\dot{a}^2},
\end{equation}
where $K = 0, \pm1$. Again we have to deal with the basic point of our
VSL cosmologies: Which $c$ are we dealing with?  We cannot simply
replace $c \to \cp$ in the above (as done in other VSL
implementations).  As we have pointed out, the $c$ appearing here must
be the fixed $\cn$, otherwise the Bianchi identities are violated and
Einstein gravity loses its geometrical interpretation in terms of
spacetime curvature. Thus we have to take $\epsilon={K
\;\cn^{2}}/{\dot{a}^2}$.  Differentiating this equation, we see that
purely on kinematic grounds
\begin{equation}
\dot{\epsilon} =
- 2K\;\cn^{2} \left({\ddot{a}}/{\dot{a}^{3}}\right)
=
-2\epsilon \left({\ddot{a}}/{\dot{a}}\right).
\end{equation}  
Given the way we have implemented VSL cosmology in terms of a
two-metric model, this equation is independent of the details 
in the photon sector. In particular, if we want to solve the flatness
problem by making $\epsilon=0$ a stable fixed point of this evolution
equation (at least for some significant portion of the history of the
universe), then we must have $\ddot a > 0$, which is equivalent to SEC
violation in FLRW.  VSL effects by themselves are not
sufficient. [Superficially similar VSL models \cite{Clayton-Moffat-1} are
claimed to solve the flatness puzzle. See \cite{BLMV99} for a
discussion of such an apparent discrepancy.]

\subsection{Monopoles and Relics}

The Kibble mechanism predicts topological defect densities that are
inversely proportional to powers of the correlation length of the
Higgs fields.  These are generally upper bounded by the Hubble
distance $c/H$.  Inflation solves this problem by diluting the density
of defects to an acceptable degree. We deal with the issue by varying
$c$ in such a way as to have a large Hubble distance during defect
formation.  Thus we need the transition in the speed of light to
happen {\em after} the SSB that leads to monopole production.
[We also want good thermal coupling between the photons and the Higgs
field, to justify using the photon horizon scale in the Kibble
freeze-out argument.] 

\section{PRIMORDIAL FLUCTUATIONS}

The inflationary framework owes its popularity not only to its ability
to strongly mitigate the main cosmological puzzles, but also to its
providing a plausible micro-physics explanation for the causal
creation of primordial perturbations.

In the case of $\chi$VSL, the creation of primordial fluctuations is
also generic. The basic mechanism can be easily understood by
modelling the change in the speed of light as due to the effect of a
changing ``effective refractive index of the EM vacuum'':
$n_{EM}={\cn}/{\cp}=
\left(\sqrt{[1+(A\,M^{-4})(\partial_{t}\chi)^2]}\right)^{-1}$.
Particle creation from a time-varying refractive index is a well-known
effect, and many features of it are similar to those derivable for its
inflationary and VSL counterparts (\emph{e.g.,} the particles are still
produced in squeezed 
pairs). We must stress the fact that these
mechanisms are not entirely identical. In $\chi$VSL cosmologies a
thermal distribution of the excited modes (with a temperature
approximately constant in time) is no longer generic, and likewise the
Harrison-Zel'dovich (HZ) spectrum is not guaranteed. Nevertheless,
approximate thermality, at fixed temperature, over a
wide frequency range can be proved for suitable regimes~\cite{BLMV99},
implying an approximate HZ spectrum of primordial perturbations only on a
finite range of frequencies. We hope that this ``weak'' prediction of a HZ
spectrum will be among the possible observational test of these
implementations of the VSL framework.

\section{CONCLUSIONS}

Implementing VSL cosmologies in a geometrically clean way seems to
lead almost inevitably to some version of a two-metric cosmology.  We
have indicated that there are several different ways of building
two-metric VSL cosmologies and have discussed some of their generic
{cosmological} features.


\end{document}